\documentstyle[12pt]{article}

\begin{document}
\sloppy
\sloppy
\sloppy
$\ $
\begin{flushright}{UT-678\\ April  '94}\end{flushright}
\vskip 1.5 truecm

\centerline{\large{\bf Generalized Pauli-Villars Regularization}}
\centerline{\large{\bf    and the Covariant Form of Anomalies}}
\vskip .75 truecm
\centerline{\bf Kazuo Fujikawa}
\vskip .4 truecm
\centerline {\it Department of Physics,University of Tokyo}
\centerline {\it Bunkyo-ku,Tokyo 113,Japan}
\vskip 1. truecm

\makeatletter
\@addtoreset{equation}{section}
\def\theequation{\thesection.\arabic{equation}}
\makeatother

\vskip 1. truecm

\begin{abstract}

\par
In the generalized Pauli-Villars regularization of chiral gauge theory
 proposed by Frolov and Slavnov ,
it is important to specify how to sum the contributions from an
infinite number of regulator fields. It is shown that an explicit sum of
contributions from an infinite number of fields in anomaly-free gauge theory
essentially results in a
specific choice of regulator in the past formulation of covariant
anomalies.
We show this correspondence by reformulating the generalized Pauli-
Villars regularization as a regularization of composite current operators.
We thus naturally understand why the covariant fermion number anomaly in the
Weinberg-Salam theory is reproduced in the generalized Pauli-Villars
regularization.
A salient feature of the covariant
regularization,which is not implemented in the lagrangian level in general but
 works for any chiral theory and gives rise to
covariant anomalies , is that it  spoils the Bose symmetry
in anomalous theory. The covariant regularization however preserves the
Bose symmetry as well as gauge invariance in anomaly-free gauge theory.
\end{abstract}

\newpage
\section{Introduction}

\par
An interesting regularization of chiral gauge theory in the Lagrangian
level has been proposed by Frolov and Slavnov \cite{1}.
This scheme incorporates an infinite number of bosonic and fermionic
regulator fields,and as such how to sum the contributions from the
infinite number of regulator fields constitutes an essential part of
this regularization ; a formal introduction of an infinite number of
regulator fields in the Lagrangian does not completely specify the theory.
Detailed analysis of this regularization scheme have been also
performed by several authors \cite{2}\cite{3} : For example ,
 the covariant form of anomaly for the fermion number current in the
Weinberg-Salam theory \cite{4} is naturally reproduced in the generalized
Pauli-Villars regularization\cite{3}.
\par
The purpose of the present paper is to analyze the basic mechanism of
this regularization scheme from a general view point of regularization
and anomalies \cite{5}. We show that the generalized Pauli-Villars
regularization , after one sums the contributions from an infinite
number of fields , essentially corresponds to a specific choice of
regulator in the calculational scheme of covariant anomalies \cite{6}
\cite{7}. The covariant form of fermion number anomaly is thus naturally
understood. A reformulation of the genaralized Pauli-Villars regularization
as a regularization of composite current operators is crucial in this
analysis. The calculational scheme of covariant anomalies , which
works for any chiral gauge theory,was introduced as a convenient means in
 the path integral formulation of anomalous identities \cite{6} \cite{8}.This
 regularization is not implemented in the Lagrangian level,but rather
 it regularizes various currents and amplitudes directly. In terms of Feynman
 diagrams , this regularization imposes the gauge invariance on all
 the vertices except for the one corresponding to the Noether current
 generated by the change of path integral variables.The anomaly
 produced at the Noether current is thus "gauge covariant" , but it
 explicitly spoils the Bose symmetry of the underlying Feynman diagrams.
 If one applies this regularization to non-anomalous diagrams , however ,
 the Noether current is conserved and the Bose symmetry in Feynman
 diagrams is preserved.
\par
In the following , we show the above correspondence ( and also some
difference ) between the generalized Pauli-Villars regularization and the
calculational scheme of covariant anomalies by explicitly evaluating
several anomalous as well as non-anomalous diagrams.

\section{Generalized Pauli-Villars  Regularization  }
\par
We first recapitulate the essence of the generalized Pauli-Villars
regularization and reformulate it as a regularization of composite
current operators. The starting theory which we want to regularize is
defined by

\begin{equation}
{\cal L}=\overline{\psi}i\not{\!\!D}\left(\frac{1+\gamma_{5}}{2}\right)\psi
\end{equation}
where
\begin{eqnarray}
\not{\!\!D}&=&\gamma^{\mu}(\partial_{\mu}-igA^{a}_{\mu}(x)T^{a})\nonumber\\
&\equiv&\gamma^{\mu}(\partial_{\mu}-igA_{\mu}(x))
\end{eqnarray}
and $T^{a}$is the hermitian generator of a compact semi-simple group,
\begin{equation}
[T^{a},T^{b}]=if^{abc}T^{c} \ \  , \ \  TrT^{a}T^{b}=\frac{1}{2}\delta^{ab}.
\end{equation}
In the main part of this paper , we treat the gauge field $A_{\mu}(x)$
 as a background field,and the dynamical aspects of $A_{\mu}$ will be
briefly commented on later. In the Euclidean metric we use , the $\gamma$
- matrices satisfy

\begin{eqnarray*}
\{\gamma^{\mu},\gamma^{\nu}\}=2g^{\mu\nu} &,&
g^{\mu\nu}=(-1,-1,-1,-1)\nonumber\\
(\gamma^{\mu})^{\dag}=-\gamma^{\mu} &,& \gamma_{5}^{\dag}=\gamma_{5}\nonumber\\
(\gamma_{5})^{2}=1.&&
\end{eqnarray*}
The Dirac operator $\not{\!\!D}$ is formally hermitian for the natural
inner product of Euclidean theory

\begin{eqnarray}
(\Phi,\not{\!\!D}\Psi)&\equiv&{\int}d^{4}x\Phi^{\dag}\not{\!\!D}\Psi\nonumber\\
&=&(\not{\!\!D}\Phi,\Psi).
\end{eqnarray}
\par
The generalized Pauli-Villars regularization of (2,1) is defined by

\begin{eqnarray}
{\cal L}&=&\overline{\psi}i\not{\!\!D}\psi-\overline{\psi}_{L}M\psi_{R}-
\overline{\psi}_{R}M^{\dag}\psi_{L}\nonumber\\
& &+\overline{\phi}i\not{\!\!D}\phi-\overline{\phi}M'\phi
\end{eqnarray}
where

\begin{eqnarray}
\psi_{R}=\frac{1}{2}(1+\gamma_{5})\psi &,& \psi_{L}=\frac{1}{2}(1-\gamma
_{5})\psi
\end{eqnarray}
and the infinite dimensional mass matrices in (2.5) are defined by

\begin{eqnarray}
M&=&\left(
\begin{array}{ccccc}
0 & 2 & 0 & 0 & \cdots \\
0 & 0 & 4 & 0 & \cdots \\
0 & 0 & 0 & 6 & \cdots \\
\cdots
\end{array}\right)\Lambda\nonumber\\
M^{\dag}M&=&\left(
\begin{array}{ccccc}
0 &   &   &   &   \\
& 2^{2} & &0 & \\
& &  4^{2} & & \\
&0 &  & 6^{2} & \\
& &  & & \ddots
\end{array}\right)\Lambda^{2}\nonumber\\
MM^{\dag}&=&\left(
\begin{array}{ccccc}
2^{2} &   &      & &   \\
& 4^{2} & &0&  \\
 & & 6^{2} & & \\
 &0 & &  \ddots&\\
 & & & &
\end{array}\right)\Lambda^{2}\nonumber\\
M'&=&\left(
\begin{array}{ccccc}
1 &   &      & &  \\
& 3 & &0 & \\
 & & 5 & & \\
 &0 & &  \ddots&\\
& & & &
\end{array}\right)\Lambda=(M')^{\dag}\nonumber\\
\end{eqnarray}
where $\Lambda$ is a parameter with dimensions of mass.
\par
The fields $\psi$ and $\phi$ in (2.5) then contain an infinite number
of components , each of which is a conventional 4-component Dirac field;$
\psi(x)$ consists of conventional anti-commuting (Grassmann) fields , and
$\phi(x)$ consists of commuting bosonic Dirac fields. The
regularization (2.5) corresponds to the so-called "vector - like"
formulation \cite{2}.
\par
The Lagrangian (2.5) is invariant under the gauge transformation
\begin{eqnarray}
\psi(x)&\rightarrow&\psi'(x)=U(x)\psi(x){\equiv}exp[iw^{a}(x)T^{a}]
\psi(x)\nonumber\\
\overline{\psi}(x)&\rightarrow&\overline{\psi}'(x)=\overline{\psi}(x)U(x)
^{\dag}\nonumber\\
\phi(x)&\rightarrow&\phi'(x)=U(x)\phi(x)\nonumber\\
\overline{\phi}(x)&\rightarrow&\overline{\phi}'(x)=
\overline{\phi}(x)U(x)^{\dag}
\nonumber\\
\not{\!\!D}&\rightarrow&\not{\!\!D}'=U(x)\not{\!\!D}U(x)^{\dag}.
\end{eqnarray}
The Noether current associated with the gauge coupling in (2.5) is
defined by the infinitesimal change of matter variables in (2.8) with
$\not{\!\!D}$ kept fixed :
\begin{eqnarray}
{\cal L}'&=&\overline{\psi}'i\not{\!\!D}\psi'-\overline{\psi}_{L}'M\psi_{R}'-
\overline{\psi}'_{R}M^{\dag}\psi'_{L}\nonumber\\
&&\ \ \ \ \
+\overline{\phi}'i\not{\!\!D}\phi'-\overline{\phi}'M'\phi'\nonumber\\
&=&-(D_{\mu}w)^{a}J^{{\mu}a}(x)+{\cal L}
\end{eqnarray}
with

\begin{equation}
J^{{\mu}a}(x)=\overline{\psi}(x)T^{a}\gamma^{\mu}\psi(x)+\overline{\phi}
(x)T^{a}\gamma^{\mu}\phi(x).
\end{equation}
Similarly , the U(1) transformation
\begin{eqnarray}
\psi(x)&\rightarrow&e^{i\alpha(x)}\psi(x)\ ,\  \overline{\psi}(x)\rightarrow
\overline{\psi}(x)e^{-i\alpha(x)}\nonumber\\
\phi(x)&\rightarrow&e^{i\alpha(x)}\phi(x)\ ,\  \overline{\phi}(x)\rightarrow
\overline{\phi}(x)e^{-i\alpha(x)}\nonumber\\
\end{eqnarray}
gives rise to the U(1) fermion number current

\begin{equation}
J^{\mu}(x)=\overline{\psi}(x)\gamma^{\mu}\psi(x)+\overline{\phi}(x)
\gamma^{\mu}\phi(x).
\end{equation}
The chiral transformation

\begin{eqnarray}
\psi(x)&\rightarrow&e^{i\alpha(x)\gamma_{5}}\psi(x)\  ,\  \overline{\psi}
\rightarrow\overline{\psi}(x)e^{i\alpha(x)\gamma_{5}}\nonumber\\
\phi(x)&\rightarrow&e^{i\alpha(x)\gamma_{5}}\phi(x)\  ,\  \overline{\phi}
\rightarrow\overline{\phi}(x)e^{i\alpha(x)\gamma_{5}}
\end{eqnarray}
gives the U(1) chiral current

\begin{equation}
J^{\mu}_{5}(x)=\overline{\psi}(x)\gamma^{\mu}\gamma_{5}\psi(x)+
\overline{\phi}(x)\gamma^{\mu}\gamma_{5}\phi(x).
\end{equation}
\par
Considering the variation of action under the transformation (2.9) and (2.11)
, one can show that the vector currents (2.10) and (2.12) are
$\underline{naively}$ conserved \footnote[1]{The fact that the regularized
currents satisfy anomaly-free relations (2.15) shows that the regularization
 (2.5)
is ineffective for the evaluation of possible anomalies in these vector
currents.}

\begin{eqnarray}
(D_{\mu}J^{\mu})^{a}(x)&\equiv&\partial_{\mu}J^{{\mu}a}(x)+gf^{abc}A_{\mu}
^{b}(x)J^{{\mu}c}(x)=0,\nonumber\\
\partial_{\mu}J^{\mu}(x)&=&0
\end{eqnarray}
whereas the chiral current (2.14) satisfies the $\underline{naive}$
identity
\begin{equation}
\partial_{\mu}J^{\mu}_{5}(x)=2i\overline{\psi}_{L}M\psi_{R}-2i\overline{\psi}
_{R}M^{\dag}\psi_{L}+2i\overline{\phi}M'\gamma_{5}\phi.
\end{equation}

\par
The quantum theory of (2.5) may be defined by the path integral as
\begin{eqnarray}
Z&=&{\int}D\overline{\psi}D{\psi}D\overline{\phi}D{\phi}exp[{\int}
{\cal L}d^{4}x]\nonumber\\
&\equiv&\int{d}{\mu}exp[{\int}{\cal L}d^{4}x]
\end{eqnarray}
and , for example ,

\begin{equation}
<\overline{\psi}(x)T^{a}\gamma^{\mu}\psi(x)>={\int}d\mu\overline{\psi}
(x)T^{a}\gamma^{\mu}\psi(x)exp[{\int}{\cal L}d^{4}x].
\end{equation}
The path integral over the bosonic variables $\phi$ and $\overline{\phi}$
for the Dirac operator in Euclidean theory needs to be defined via a
suitable rotation in the functional space.

\bigskip
\bigskip
\newpage
\begin{flushleft}
\underline{\bf{Definition of Currents in Terms of Propagators}}
\end{flushleft}
\par
We now define the currents in terms of propagators.
The basic idea of this approach is explained for the $\underline{un-
regularized}$ theory in (1) as follows : We start with the current
associated with the gauge coupling

\begin{eqnarray}
\lefteqn{<\overline{\psi}(x)T^{a}\gamma^{\mu}(\frac{1+\gamma_{5}}{2})
\psi(x)>}\nonumber\\
&=&\lim_{y{\rightarrow}x}<T^{*}\overline{\psi}(y)T^{a}\gamma^{\mu}
(\frac{1+\gamma_{5}}{2})\psi(x)>\nonumber\\
&=&-\lim_{y{\rightarrow}x}<T^{*}(T^{a})_{bc}\gamma^{\mu}_{{\alpha}\delta}
(\frac{1+\gamma_{5}}{2})_{\delta\beta}\psi_{{\beta}c}(x)\overline{\psi}
_{{\alpha}b}(y)>\nonumber\\
&=&\lim_{y{\rightarrow}x}Tr[T^{a}\gamma^{\mu}(\frac{1+\gamma_{5}}{2})
\frac{1}{i\not{\!\!D}}\delta(x-y)]
\end{eqnarray}
where we used the anti-commuting property of $\psi$ and the expression of
 the propagator

\begin{equation}
<T^{*}\psi(x)\overline{\psi}(y)>=(\frac{1+\gamma_{5}}{2})\frac{(-1)}
{i\not{\!\!D}_{x}}\delta(x-y)
\end{equation}
The trace in (2.19) runs over the Dirac and Yang-Mills indices.
We now notice the expansion

\begin{eqnarray}
\frac{1}{i\not{\!\!D}}&=&\frac{1}{i\not{\!\partial}+g\not{\!\!A}}\nonumber\\
&=&\frac{1}{i\not{\!\partial}}+\frac{1}{i\not{\!\partial}}(-g
\not{\!\!A})\frac{1}{i\not{\!\partial}}\nonumber\\
&&+\frac{1}{i\not{\!\partial}}(-g
\not{\!\!A})\frac{1}{i\not{\!\partial}}(-g\not{\!\!A})
\frac{1}{i\not{\!\partial}}+\cdots
\end{eqnarray}
When one inserts (2.21) into (2.19) and retains only the terms linear in
$A^{b}_{\nu}(x)$ , one obtains

\begin{eqnarray}
\lefteqn{\lim_{y{\rightarrow}x}Tr[T^{a}\gamma^{\mu}(\frac{1+\gamma_{5}}{2})
\frac{(-1)}{i\not{\!\partial}}\gamma^{\nu}T^{b}gA^{b}_{\nu}(x)
\frac{1}{i\not{\!\partial}}\delta(x-y)]}\nonumber\\
&=&\lim_{y{\rightarrow}x}{\int}d^{4}zTr[T^{a}\gamma^{\mu}(\frac{1+
\gamma_{5}}{2})\frac{(-1)}{i\not{\!\partial}}\nonumber\\
&& \ \ \ {\times}\delta(x-z)T^{b}\gamma^{\nu}
\frac{1}{i\not{\!\partial}}\delta(x-y)]gA^{b}_{\nu}(z)
\end{eqnarray}
where the derivative $\partial_{\mu}$ acts on $\underline{all}$ the x-
variables standing on the right of it in (2.22).
If one takes the variational derivative of (2.22) with respect to $gA
^{b}_{\nu}(z)$ , one obtains

\begin{eqnarray}
\lefteqn{\lim_{y{\rightarrow}x}Tr[T^{a}\gamma^{\mu}(\frac{1+\gamma_{5}}{2})
\frac{(-1)}{i\not{\!\partial}}\delta(x-z)\gamma^{\nu}T^{b}
\frac{1}{i\not{\!\partial}}\delta(x-y)]}\nonumber\\
&=&\lim_{y{\rightarrow}x}\int\frac{d^{4}q}{(2\pi)^{4}}\frac{d^{4}k}
{(2\pi)^{4}}Tr[T^{a}
\gamma^{\mu}(\frac{1+\gamma_{5}}{2})\frac{(-1)}{\not{\!k}+\not{\!q}}
T^{b}\gamma^{\nu}\frac{1}{\not{\!k}}]e^{-iq(x-z)}e^{-ik(x-y)}\nonumber\\
&=&\int\frac{d^{4}q}{(2\pi)^{4}}e^{-iq(x-z)}(-1)
\int\frac{d^{4}k}{(2\pi)^{4}}Tr[T^{a}
\gamma^{\mu}(\frac{1+\gamma_{5}}{2})\frac{1}{\not{\!k}+\not{\!q}}
T^{b}\gamma^{\nu}(\frac{1+\gamma_{5}}{2})\frac{1}{\not{\!k}}]\nonumber\\
&\equiv&\int\frac{d^{4}q}{(2\pi)^{4}}e^{-iq(x-z)}\Pi^{ab}_{\mu\nu}(q)
\end{eqnarray}
where we used the representations of $\delta$-function

\begin{eqnarray}
\delta(x-z)&=&\int\frac{d^{4}q}{(2\pi)^{4}}e^{-iq(x-z)}\nonumber\\
\delta(x-y)&=&\int\frac{d^{4}k}{(2\pi)^{4}}e^{-ik(x-y)}.
\end{eqnarray}
\par
The last expression in (2.23) stands for the vacuum polarization tensor.
Namely , one can generate the multiple correlation functions of currents
$\overline{\psi}T^{a}\gamma^{\mu}(\frac{1+\gamma_{5}}{2})\psi$ in the
perturbative sense by taking the variational derivative of (2.19) with
respect to gauge fields $A_{\mu}^{a}$ .
This idea also works for the non-gauge currents (2.12) and (2.14).
We emphasize that we always take the limit $y=x$ first before the
explicit calculation , and thus (2.19) \underline{differs}
 from the point-splitting
definition of currents.
\par
We now generalize the above definition of currents for the theory
defined by (2.5).
For this purpose , we rewrite (2.5) as
\begin{equation}
{\cal L}=\overline{\psi}i{\cal D}\psi+\overline{\phi}i{\cal D}'\phi
\end{equation}
with
\begin{eqnarray}
\cal D&\equiv&\not{\!\!D}+iM(\frac{1+\gamma_{5}}{2})+iM^{\dag}
(\frac{1-\gamma_{5}}{2})\nonumber\\
\cal D'&\equiv&\not{\!\!D}+iM'.
\end{eqnarray}
The gauge current (2.10) is then defined by

\begin{eqnarray}
J^{{\mu}a}(x)&=&\lim_{y{\rightarrow}x}\{<T^{*}\overline{\psi}(y)T^{a}
\gamma^{\mu}\psi(x)>+<T^{*}\overline{\phi}(y)T^{a}
\gamma^{\mu}\phi(x)>\}\nonumber\\
&=&\lim_{y{\rightarrow}x}\{-<T^{*}T^{a}\gamma^{\mu}\psi(x)\overline{\psi}
(y)>+<T^{*}T^{a}\gamma^{\mu}\phi(x)\overline{\phi}(y)>\}\nonumber\\
&=&\lim_{y{\rightarrow}x}Tr[T^{a}\gamma^{\mu}(\frac{1}{i\cal D}-\frac{1}{i\cal
D'})
\delta(x-y)]
\end{eqnarray}
where trace includes the sum over the infinite number of field components
 in addition to Dirac and Yang-Mills indices.
 The anti-commuting property of $\psi(x)$ and the commuting property of $
\phi(x)$ are used in (2.27).
\par
We next notice the relations

\begin{eqnarray}
\frac{1}{\cal D}&=&\frac{1}{\cal D^{\dag}\cal D}\cal D^{\dag}\nonumber\\
&=&\frac{1}{\not{\!\!D}^{2}+\frac{1}{2}M^{\dag}M(1+\gamma_{5})+\frac{1}{2}
MM^{\dag}(1-\gamma_{5})}\cal D^{\dag}\nonumber\\
&=&[(\frac{1+\gamma_{5}}{2})\frac{1}{\not{\!\!D}^{2}+M^{\dag}M}+(\frac{
1-\gamma_{5}}{2})\frac{1}{\not{\!\!D}^{2}+MM^{\dag}}]\nonumber\\
&&\ \ \ \ \ \times[\not{\!\!D}-iM^{\dag}(\frac{
1+\gamma_{5}}{2})-iM(\frac{1-\gamma_{5}}{2})]\nonumber\\
\frac{1}{\cal D'}&=&\frac{1}{(\cal D')^{\dag}\cal D'}(\cal
D')^{\dag}\nonumber\\
&=&\frac{1}{\not{\!\!D}^{2}+(M')^{2}}(\not{\!\!D}-iM').
\end{eqnarray}
We thus rewrite (2.27) as

\begin{eqnarray}
\lefteqn{Tr\left[-iT^{a}\gamma^{\mu}(\frac{1}{\cal D}-\frac{1}{\cal D'})
\delta(x-y)\right]}
\nonumber\\
&=&Tr\left\{-iT^{a}\gamma^{\mu}\left[(\frac{1+\gamma_{5}}{2})\sum^{\infty}
_{n=0}
\frac{1}{\not{\!\!D}^{2}+(2n\Lambda)^{2}}\right.\right.\nonumber\\
&&\ \ \ \ \ \ \ \ +(\frac{1-\gamma_{5}}{2})
\sum^{\infty}_{n=1}\frac{1}{\not{\!\!D}^{2}+(2n\Lambda)^{2}}\nonumber\\
& &\ \ \ \ \ \ \ \  \left.\left.-\sum^{\infty}_{n=0}\frac{1}{\not{\!\!D}
^{2}+[(2n+1)\Lambda]^{2}}\right]
\not{\!\!D}\delta(x-y)\right\}\nonumber\\
&=&\frac{1}{2}Tr\left[-iT^{a}\gamma^{\mu}\sum^{\infty}_{n=-\infty}
\frac{(-1)^{n}\not{\!\!D}^{2}}{\not{\!\!D}^{2}+(n\Lambda)^{2}}
\frac{1}{\not{\!\!D}}\delta(x-y)\right]\nonumber\\
& &\ \ \ \ \ \ \ \ +\frac{1}{2}Tr\left[-iT^{a}\gamma^{\mu}\gamma_{5}
\frac{1}{\not{\!\!D}}
\delta(x-y)\right]\nonumber\\
&=&\frac{1}{2}Tr\left[T^{a}\gamma^{\mu}f(\not{\!\!D}^{2}/\Lambda^{2})
\frac{1}{i\not{\!\!D}}\delta(x-y)\right]\nonumber\\
& &\ \ \ \ \ \ \ \ +\frac{1}{2}Tr\left[T^{a}\gamma^{\mu}
\gamma_{5}\frac{1}{i\not{\!\!D}}
\delta(x-y)\right]
\end{eqnarray}
where we explicitly evaluated the trace over the infinite number of components
and used the fact that the trace over an odd number of $\gamma$-matrices
 vanishes.
We also defined $f(x^{2})$ by

\begin{eqnarray}
f(x^{2})&\equiv&\sum^{\infty}_{n=-\infty}\frac{(-1)^{n}x^{2}}{x^{2}+
(n\Lambda)^{2}}\nonumber\\
&=&\frac{(\pi{x}/\Lambda)}{sinh({\pi}x/\Lambda)}.
\end{eqnarray}
This last expression of (2.30) as a sum of infinite number of terms is
given in ref.\cite{1}.
The regulator $f(x^{2})$, which rapidly approaches 0 at $x^{2}=\infty$,
satisfies

\begin{eqnarray}
f(0)&=&1\nonumber\\
x^{2}f'(x^{2})&=&0 \ for\   x\rightarrow{0}\nonumber\\
f(+\infty)&=&f'(+\infty)=f''(+\infty)=\cdots=0\nonumber\\
x^{2}f'(x^{2})&\rightarrow&0 \ \  for \ \  x\rightarrow\infty.
\end{eqnarray}
\par
The essence of the generalized Pauli-Villars regularization (2.5) is thus
 summarized in terms of regularized currents as follows:

\begin{eqnarray}
\lefteqn{<\overline{\psi}(x)T^{a}\gamma^{\mu}(\frac{1+\gamma_{5}}{2})\psi(x)>
_{PV}}\nonumber\\
&=&\lim_{y{\rightarrow}x}\left\{\frac{1}{2}Tr\left[T^{a}
\gamma^{\mu}f(\not{\!\!D}^{2}
/\Lambda^{2})\frac{1}{i\not{\!\!D}}\delta(x-y)\right]\right.\nonumber\\
& &\ \
\left.+\frac{1}{2}Tr\left[T^{a}\gamma^{\mu}\gamma_{5}
\frac{1}{i\not{\!\!D}}\delta
(x-y)\right]\right\}\nonumber\\
\lefteqn{<\overline{\psi}(x)\gamma^{\mu}(\frac{1+\gamma_{5}}{2})\psi(x)>_{PV}}
\nonumber\\
&=&\lim_{y{\rightarrow}x}\left\{\frac{1}{2}Tr\left[\gamma^{\mu}
f(\not{\!\!D}^{2}/\Lambda^{2})
\frac{1}{i\not{\!\!D}}\delta(x-y)\right]\right.\nonumber\\
& &\ \ \left.+\frac{1}{2}Tr\left[\gamma^{\mu}\gamma_{5}\frac{1}{i\not{\!\!D}}
\delta(x-y)\right]\right\}\nonumber\\
\lefteqn{<\overline{\psi}(x)\gamma^{\mu}\gamma_{5}(\frac{1+\gamma_{5}}{2})
\psi(x)>_{PV}}\nonumber\\
&=&\lim_{y{\rightarrow}x}\left\{\frac{1}{2}Tr\left[
\gamma^{\mu}\gamma_{5}f(\not{\!\!D}^{2}/\Lambda^{2})
\frac{1}{i\not{\!\!D}}\delta(x-y)\right]\right.\nonumber\\
& &\ \ \left.+\frac{1}{2}Tr\left[\gamma^{\mu}\frac{1}
{i\not{\!\!D}}\delta(x-y)\right]\right\}.
\end{eqnarray}
In the left-hand sides of (2.32), the currents are defined in terms of
the original fields appearing in (2.1).
The axial-vector and vector $U(1)$ currents written in terms of the original
fields in (2.1) are identical , but the regularized versions (i.e. the
last two equations in (2.32)) are different. In particular , the vector $U(1)$
 current(i.e. , the second equation in (2.32)) is not completely regularized.
See also refs.\cite{2} and \cite{3}. This reflects the different form of
\underline{naive} identities in (2.15) and (2.16) ; if all the currents
are well regularized , the \underline{naive} form of identities would
also coincide. We emphasize that all the one-loop diagrams are generated
 from the (partially) regularized currents in (2.32) ; in other words ,
(2.32) retains all the information of the generalized Pauli-Villars
regularization (2.5).

\par
The trace of energy-momentum tensor generated by the matter field in (2.1)
 is also interesting.
This is related to the variation of field variables\cite{8}

\begin{equation}
\psi(x)\rightarrow{e}^{-\frac{1}{2}\alpha(x)}\psi(x)\ ,\ \overline{\psi}(x)
\rightarrow\overline{\psi}(x)e^{-\frac{1}{2}\alpha(x)}
\end{equation}
which is the flat space-time limit of the variation of the weighted
variables

\begin{equation}
\tilde{\psi}(x)\equiv{g}^{\frac{1}{4}}\psi(x)\ ,\  \tilde{\overline{\psi}}
(x)\equiv{g}^{\frac{1}{4}}\overline{\psi}(x)
\end{equation}
under the Weyl transformation

\begin{eqnarray}
\psi(x)&{\rightarrow}&e^{\frac{3}{2}\alpha(x)}\psi(x)\ \ ,\ \
\overline{\psi}(x){\rightarrow}\overline{\psi}(x)
e^{\frac{3}{2}\alpha(x)}\nonumber\\
g_{\mu\nu}(x)&\rightarrow&e^{-2\alpha(x)}g_{\mu\nu}(x)\ \ ,\ \
g=detg_{\mu\nu}.\nonumber\\
\end{eqnarray}
The Noether density generated by the (infinitesimal) transformation (2.33)
 in (2.1) is given by

\begin{eqnarray}
\lefteqn{\int{d}^{4}x\overline{\psi}'(x)i\not{\!\!D}
(\frac{1+\gamma_{5}}{2})\psi'
(x)}\nonumber\\
&=&-\int{d}^{4}x\alpha(x){\overline{\psi}(x)\frac{i}{2}\stackrel
{\leftrightarrow}
{\not{\!\!D}}
(\frac{1+\gamma_{5}}{2})\psi(x)}\nonumber\\
&&\ \ \ \ \ +\int{d}^{4}x\overline{\psi}(x)i
\not{\!\!D}(\frac{1+\gamma_{5}}{2})\psi(x)
\end{eqnarray}.
\par
Following the same procedure as in (2.32), we find

\begin{eqnarray}
\lefteqn{<\overline{\psi}(x)\frac{i}{2}\stackrel{\leftrightarrow}{\not{\!\!D}}
(\frac{1+\gamma_{5}}{2})\psi(x)>
_{PV}} \nonumber\\
&=&\lim_{y{\rightarrow}x}(\frac{i}{2})\{<T^{*}\overline{\psi}(y)
\not{\!\!D}_{x}(\frac{1+\gamma_{5}}{2})\psi(x)>_{PV}\nonumber\\
&& \ \ \ \ \ -<T^{*}
\overline{\psi}(y)\stackrel{\leftarrow}{\not{\!\!D}_{y}}
(\frac{1+\gamma_{5}}{2})\psi(x)>
_{PV}\}.\nonumber\\
&=&\lim_{y{\rightarrow}x}(\frac{1}{2})Tr\left[f(\not{\!\!D}^{2}/\Lambda^{2})
\delta(x-y)\right].
\end{eqnarray}

\par
In summary, eqs (2.30),(2.32) and (2.37) are the basic results of the
generalized Pauli-Villars regularization (2.5).

\section{Covariant Regularization  and Covariant Anomalies}

The calculational scheme of covariant anomalies starts with regularized
current operators \cite{9} for the theory in (2.1) as follows:

\begin{eqnarray}
\lefteqn{<\overline{\psi}(x)T^{a}\gamma^{\mu}(\frac{1+\gamma_{5}}{2})\psi(x)>
_{cov}}\nonumber\\
&=&\lim_{y{\rightarrow}x}Tr\left[T^{a}\gamma^{\mu}(\frac{1+\gamma_{5}}{2})
f(\not{\!\!D}^{2}/\Lambda^{2})\frac{1}{i\not{\!\!D}}\delta(x-y)\right]
\nonumber\\
&=&\sum_{n}\phi_{n}(x)^{\dag}\left[T^{a}\gamma^{\mu}(\frac{1+\gamma_{5}}{2})f(
\lambda_{n}^{2}/\Lambda^{2})\frac{1}{i\lambda_{n}}\right]\phi_{n}(x)\nonumber\\
\lefteqn{<\overline{\psi}(x)\gamma^{\mu}(\frac{1+\gamma_{5}}{2})\psi(x)>
_{cov}}\nonumber\\
&=&\lim_{y{\rightarrow}x}Tr\left[\gamma^{\mu}(\frac{1+\gamma_{5}}{2})
f(\not{\!\!D}^{2}/\Lambda^{2})\frac{1}{i\not{\!\!D}}\delta(x-y)
\right]\nonumber\\
&=&\sum_{n}\phi_{n}(x)^{\dag}\left[\gamma^{\mu}(\frac{1+\gamma_{5}}{2})f(
\lambda_{n}^{2}/\Lambda^{2})\frac{1}{i\lambda_{n}}\right]\phi_{n}(x)\nonumber\\
\lefteqn{<\overline{\psi}(x)\frac{i}{2}\stackrel{\leftrightarrow}{\not{\!\!D}}
(\frac{1+\gamma_{5}}{2})\psi(x)>
_{cov}}\nonumber\\
&=&\lim_{y{\rightarrow}x}Tr\left[(\frac{1}{2})
f(\not{\!\!D}^{2}/\Lambda^{2})\delta(x-y)\right]\nonumber\\
&=&\frac{1}{2}\sum_{n}\phi_{n}(x)^{\dag}f(\lambda_{n}^{2}/\Lambda^{2})
\phi_{n}(x)
\end{eqnarray}
where the complete set $\{\phi_{n}(x)\}$ is defined by

\begin{eqnarray}
\not{\!\!D}\phi_{n}(x)&\equiv&\lambda_{n}\phi_{n}(x)\nonumber\\
\int\phi_{m}(x)^{\dag}\phi_{n}(x)d^{4}x&=&\delta_{m,n}\nonumber\\
\delta_{\alpha\beta}\delta(x-y)&\rightarrow&\sum_{n}\phi_{n}(x)_{\alpha}
\phi_{n}(y)^{\dag}_{\beta}
\end{eqnarray}
with $\alpha$ and $\beta$ including Dirac and Yang-Mills indices.
\par
The function $f(x^{2})$ in (3.1) is $\underline{any}$ smooth function
which satisfies the condition (2.31).
For the  moment, we assume that there is no zero eigenvalue in (3.2).
One recognizes a close relation between the generalized Pauli-Villars
regularization ( (2.32) and (2.37) ) and the present covariant
calculational scheme.
The characteristic feature of (3.1) is that it treats the vector and
axial-vector components on an equal footing and regularizes them
simultaneously. In principle , one could apply different regulator functions ,
for example , $f(\not{\!\!\!D}^{2}/\Lambda^{2})$ and
$g(\not{\!\!\!D}^{2}/\Lambda^{2})$ respectively to vector and axial-vector
components in (3.1) instead of using  $f(\not{\!\!D}^{2}/\Lambda^{2})$ for
both of them. In this case , (2.32) is obtained as a special case of (3.1)
by taking the limit of either  $g(\not{\!\!D}^{2}/\Lambda^{2})=1$ or
 $f(\not{\!\!D}^{2}/\Lambda^{2})=1$. In this limit , however , not all the
currents are completely regularized.

\par
The anomaly for the current in (3.1) is evaluated as

\begin{eqnarray}
\lefteqn{D_{\mu}<\overline{\psi}(x)T^{a}\gamma^{\mu}(\frac{1+\gamma_{5}}{2})
\psi(x)>_{cov}}\nonumber\\
&\equiv&\partial_{\mu}<\overline{\psi}(x)T^{a}\gamma^{\mu}
(\frac{1+\gamma_{5}}{2})
\psi(x)>_{cov}+gf^{abc}A^{b}_{\mu}<\overline{\psi}(x)T^{c}\gamma^{\mu}
(\frac{1+\gamma_{5}}{2})\psi(x)>_{cov}\nonumber\\
&=&\sum_{n}\left[\phi_{n}(x)^{\dag}T^{a}
(\frac{1-\gamma_{5}}{2})f(\lambda^{2}_{n}
/\Lambda^{2})\frac{1}{i\lambda_{n}}(\not{\!\!D}\phi_{n}(x))\right.\nonumber\\
&&\ \ \ \ \
\left.-(\not{\!\!D}\phi_{n}(x))^{\dag}T^{a}(\frac{1+\gamma_{5}}{2})
f(\lambda^{2}_{n}
/\Lambda^{2})\frac{1}{i\lambda_{n}}\phi_{n}(x)\right]\nonumber\\
&=&\sum_{n}(-i)\phi_{n}(x)^{\dag}T^{a}\left[(\frac{1-\gamma_{5}}{2})
-(\frac{1+\gamma_{5}}{2})\right]f(\lambda^{2}_{n}/
\Lambda^{2})\phi_{n}(x)\nonumber\\
&=&i\sum_{n}\phi_{n}(x)^{\dag}T^{a}\gamma_{5}f(\lambda^{2}_{n}/\Lambda^{2})
\phi_{n}(x)\nonumber\\
&=&i\sum_{n}\phi_{n}(x)^{\dag}T^{a}\gamma_{5}f(\not{\!\!D}^{2}/\Lambda^{2})
\phi_{n}(x)\nonumber\\
&=&iTr\int\frac{d^{4}k}{(2\pi)^{4}}e^{-ikx}T^{a}\gamma_{5}f(\not{\!\!D}
^{2}/\Lambda^{2})e^{ikx}\nonumber\\
&=&(\frac{ig^{2}}{32\pi^{2}})TrT^{a}\epsilon^{\mu\nu\alpha\beta}F_{\mu\nu}
F_{\alpha\beta}\ \  for\ \  \Lambda\rightarrow\infty
\end{eqnarray}
where we used the relation (3.2).
We also normalized the anti-symmetric symbol as

\begin{equation}
\epsilon^{1230}=\epsilon^{1234}=1.
\end{equation}
In the last step of the calculation in (3.3), we replaced the complete
set $\{\phi_{n}(x)\}$ by the plane wave basis for the well-defined operator
$f(\not{\!\!D}^{2}/\Lambda^{2})$.
 The final result in (3.3) with $F_{\mu\nu}=(\partial_{\mu}A^{a}_{\nu}
-\partial_{\nu}A^{a}_{\mu}+gf^{abc}A^{b}_{\mu}A^{c}_{\nu})T^{a}$
holds for $\underline{any}$ function $f(x^{2})$ which satisfies (2.31).
 This fact is explained in Appendix for the sake of completeness.
Eq.(3.3) shows that only the axial-vector component contributes to the
anomaly.
\par
Similarly, the U(1) current in (3.1) satisfies the identity

\begin{eqnarray}
\lefteqn{\partial_{\mu}<\overline{\psi}(x)\gamma^{\mu}(\frac{1+\gamma_{5}}{2})
\psi(x)>_{cov}}\nonumber\\
&=&\sum_{n}\left[-(\not{\!\!D}\phi_{n}(x))^{\dag}(\frac{1+\gamma_{5}}{2})
f(\lambda_{n}^{2}/\Lambda^{2})\frac{1}{i\lambda_{n}}\phi_{n}(x)\right.
\nonumber\\
& &\ \ \ \ \ \left.+\phi_{n}(x)^{\dag}(\frac{1-\gamma_{5}}{2})
f(\lambda_{n}^{2}/\Lambda^{2})\frac{1}{i\lambda_{n}}(\not{\!\!D}\phi_{n}
(x))\right]\nonumber\\
&=&i\sum_{n}\phi_{n}(x)^{\dag}\gamma_{5}f(\lambda^{2}_{n}/\Lambda^{2})\phi
_{n}(x)\nonumber\\
&=&iTr\int\frac{d^{4}k}{(2\pi)^{4}}e^{-ikx}\gamma_{5}f(\not{\!\!D}^{2}/
\Lambda^{2})e^{ikx}\nonumber\\
&=&(\frac{ig^{2}}{32\pi^{2}})Tr\epsilon^{\mu\nu\alpha\beta}F_{\mu\nu}
F_{\alpha\beta} \ \  for\ \   \Lambda\rightarrow\infty
\end{eqnarray}
which is the result used in the analysis of baryon number violation
\cite{4} and naturally agrees with the result on the basis of the last current
 in (2.32) in the generalized Pauli-Villars regularization \cite{3}.
Again, only the axial component contributes to the anomaly.
\par
The Weyl anomaly in the last relation in (3.1) is evaluated as

\begin{eqnarray}
<\overline{\psi}(x)\frac{i}{2}\stackrel{\leftrightarrow}{\not{\!\!D}}
(\frac{1+\gamma_{5}}{2})\psi(x)
>_{cov}&=&\frac{1}{2}Tr\int\frac{d^{4}k}{(2\pi)^{4}}e^{-ikx}f(\not{\!\!
D}^{2}/\Lambda^{2})e^{ikx}\nonumber\\
&=&\frac{1}{2}(\frac{g^{2}}{24\pi^{2}})TrF^{\mu\nu}F_{\mu\nu}\ \  for\ \
\Lambda\rightarrow\infty
\end{eqnarray}
which is also known to be independent of the choice of $f(x^{2})$ in
(2.31)\cite{8}.
The coefficient of $TrF^{\mu\nu}F_{\mu\nu}$ in (3.6) gives the lowest
order fermion contribution to the renormalization group $\beta$-function
; $\beta(g_{r})=(\frac{1}{2})g^{3}_{r}/(24\pi^{2})$.
\par
The anomaly in (3.3) is covariant under gauge transformation , which is
the reason why (3.3) is called "covariant anomaly" \cite{10}.
{}From the diagramatic view point , all the vertices of one-loop diagrams
are regularized by the gauge invariant regulator $f(\lambda_{n}^{2}/
\Lambda^{2}) \ \  \underline{except}$ for the vertex corresponding to the
Noether current itself.\footnote[1]{In the generalized Pauli-Villars
regularization , all the vertices are treated on an equal footing and Bose
symmetrically since the regularization is implemented in the Lagrangian
level. The evaluation of anomaly for gauge couplings , if it should be
performed in the generalized Pauli-Villars regularization , would
therefore be quite different from the anomaly calculation in the covariant
regularization (3.1). The axial-vector component of gauge current ,
which could produce gauge anomaly , is however not regularized in (2.5)
, as  is seen in (2.32). The regularized gauge current satisfies the naive
identity (2.15) and , consequently , the regularization (2.5)  as it stands
 is not applicable to the evaluation of possible gauge anomaly.}
Because of this asymmetric treatment of vertices, the anomaly (3.3)
does not satisfy the so-called integrability (or Wess-Zumino consistency)
 condition \cite{11}.
The relation (3.3) however specifies precisely the essence of the anomaly,
namely, one $\underline{cannot}$ impose gauge invariance on all the
 vertices of anomalous diagrams.
It is also known that one can readily convert the covariant anomaly in (3.
3) to the anomaly which satisfies the Wess-Zumino condition \cite{10}.
\par
The anomaly in (3.3) vanishes for

\begin{equation}
TrT^{a}\{T^{b},T^{c}\}=0.
\end{equation}
For the anomaly-free gauge theory, which satisfies (3.7), one can
impose gauge invariance on all the gauge vertices, and consequently the
Bose symmetry is recovered.
The regularization (3.1) thus provides a natural regularization of all
the one-loop diagrams in anomaly-free gauge theory.
\par
{}From the analysis presented above, one can understand the consistency of
the "partial regularization" of the generalized Pauli-Villars
regularization (2.32) and (2.37) for anomaly-free gauge theory, except
for the second expression in (2.32) which cannot produce U(1) anomaly
by the $\underline{naive}$ treatment of the right-hand side.
Our analysis, which is based on well-regularized operators in (3.1),
provides a transparent way to understand the conclusions in refs.\cite{1}
,\cite{2} and \cite{3}.
\par
We now add several comments on the covariant regularization (3.1).
First of all, the regularization (3.1) should not be confused with the
higher derivative regularization, for example ,
\begin{equation}
{\cal L}=\overline{\psi}(x)i\not{\!\!D}(\frac{\not{\!\!D}^{2}+\Lambda^{2}}{
\Lambda^{2}})^{2}(\frac{1+\gamma_{5}}{2})\psi(x).
\end{equation}
If one chooses $f(x^{2})\equiv\left[\Lambda^{2}/(x^{2}+\Lambda^{2})\right]^{2}$
in (3.1), the regularization (3.1) resembles the theory defined by (3.8).
However, this resemblance is spurious, since the Noether current given by
 (3.8) contains higher derivative terms in addition to the minimal one
and thus the one-loop diagrams are $\underline{not} $ regularized by (3.8).
It is an interesting question whether (3.1) for anomaly-free gauge theory
 can be implemented in the Lagrangian level if one incorporates an
 infinite number of regulator fields.
\par
Secondly, the regularization (3.1) can be implemented for more general
theory such as

\begin{eqnarray}
{\cal L}&=&\overline{\psi}i\gamma^{\mu}\left[\partial
_{\mu}-iR^{a}_{\mu}(x)T^{a}
(\frac{1+\gamma_{5}}{2})-iL^{a}_{\mu}(x)T^{a}(\frac{1-\gamma_{5}}{2})\right]
\psi\nonumber\\
&=&\overline{\psi}i\gamma^{\mu}(\partial_{\mu}-iR_{\mu})
(\frac{1+\gamma_{5}}{2})\psi+\overline{\psi}i\gamma^{\mu}(\partial_{\mu}
-iL_{\mu})(\frac{1-\gamma_{5}}{2})\psi.
\end{eqnarray}
In this case, left-and right-handed components separately satisfy the
identities\cite{5,9}

\begin{equation}
D_{\mu}<\overline{\psi}T^{a}\gamma^{\mu}(\frac{1\pm\gamma{5}}{2})
\psi>_{cov}=\pm
(\frac{i}{32\pi^{2}})TrT^{a}\epsilon^{\mu\nu\alpha\beta}F_{\mu\nu}
F_{\alpha\beta}.
\end{equation}
\par
The covariant regularization is quite flexible and works for arbitrary
gauge theory ; for example, one can readily show that Yukawa couplings do
 not modify anomaly \cite{9}.
\par
Finally, we comment on the treatment of zero modes of $\not{\!\!\!D}$ in
(3.1) in some detail ; this problem is also shared by the generalized
Pauli-Villars regularization (2.32) in non-perturbative analysis.
When the gauge field $A_{\mu}(x)$ is topologically non-trivial, the
eigenvalue equation (3.2) contains well-defined zero eigenvalues \cite{12}.
The definition of current operators thus becomes subtle, but the
divergence of currents such as (3.3) and (3.5) does not contain the
singular factor $1/\lambda_{n}$ and thus anomalies themselves
are well-defined.
In the path integral framework, this situation is treated in the
following way:
\par
One first notices that $\gamma_{5}\phi_{n}(x)$ belongs to the eigenvalue
$-\lambda_{n}$ in (3.2) since $\{\gamma_{5},\not{\!\!D}\}=0$.
We thus define new complete basis sets \cite{6}

\begin{eqnarray}
\phi^{R}_{n}(x)&\equiv&(\frac{1+\gamma_{5}}{\sqrt{2}})\phi_{n}(x)
\ \  if \  \lambda_{n}>0\nonumber\\
&\equiv&(\frac{1+\gamma_{5}}{2})\phi_{n}(x) \ \ if \  \lambda_{n}=0\nonumber\\
\phi^{L}_{n}(x)&\equiv&(\frac{1-\gamma_{5}}{\sqrt{2}})\phi_{n}(x)
\ \ if \  \lambda_{n}>0\nonumber\\
&\equiv&(\frac{1-\gamma_{5}}{2})\phi_{n}(x)\ \  if \  \lambda_{n}=0.
\end{eqnarray}
Note that $\phi_{n}(x)$ with $\lambda_{n}=0$ can be chosen to be the
eigenvector of $\gamma_{5}$.
We thus expand
\begin{eqnarray}
\psi_{R}(x)&=&(\frac{1+\gamma_{5}}{2})\psi(x)\nonumber\\
&=&\sum_{\lambda_{n}\geq0}a_{n}\phi_{n}^{R}(x)\nonumber\\
\overline{\psi}_{R}(x)&=&\sum_{\lambda_{n}\geq0}\overline{b_{n}}\phi^{L}
_{n}(x)^{\dag}
\end{eqnarray}
where $a_{n}$ and $\overline{b_{n}}$ are Grassmann numbers, and the
action (2.1) and the path integral measure are formally defined by

\begin{eqnarray}
S&=&\int{{\cal L}}d^{4}x=\sum_{\lambda_{n}>0}\lambda_{n}\overline{b_{n}}a_{n}
\nonumber\\
d\mu&=&{\cal D}\overline{\psi}_{R}{\cal D}\psi_{R}=\prod_{\lambda_{n}\geq0}
d\overline{b_{n}}da_{n}
\end{eqnarray}
The Jacobian factor under the (infinitesimal) chiral U(1) transformation

\begin{eqnarray}
\psi_{R}'(x)&=&e^{i\alpha(x)\gamma_{5}}\psi_{R}(x)=e^{i\alpha(x)}
\psi_{R}(x)\nonumber\\
\overline{\psi}_{R}'(x)&=&\overline{\psi}_{R}(x)e^{-i\alpha(x)}
\end{eqnarray}
is given by

\begin{eqnarray}
J&=&exp\left[-i\int{d}^{4}x\alpha(x)\sum_{\lambda_{n}\geq0}
\{\phi_{n}^{R}(x)^{\dag}
\phi^{R}_{n}(x)-\phi_{n}^{L}(x)^{\dag}\phi^{L}_{n}(x)\}\right]\nonumber\\
&=&exp\left[-i\int{d}^{4}x\alpha(x)\sum_{all\lambda_{n}}\phi_{n}^{\dag}(x)
\gamma_{5}\phi_{n}(x)\right]
\end{eqnarray}

\par
The sum of terms in (3.15) may be defined by
\begin{equation}
\lim_{N{\rightarrow}\infty}\sum^{N}_{n=1}\phi_{n}(x)^{\dag}\gamma_{5}
\phi_{n}(x)=\lim_{\Lambda\rightarrow\infty}\sum^{\infty}_{n=1}\phi
_{n}(x)^{\dag}\gamma_{5}f(\lambda_{n}^{2}/\Lambda^{2})\phi_{n}(x)
\end{equation}
by replacing the mode cut-off, which is natural for the definition (3.13),
 by the cut-off in $\lambda_{n}$ by using a suitable regulator satisfying
 (2.31).
By this way, one directly obtains the anomaly factor (3.5) as a
Jacobian without referring to current operators.
In the generalized Pauli-Villars regularization (2.17) the measure is
shown to be
invariant under the chiral U(1) transformation\footnote[1]{As for
 the vector-like transformation corresponding to the naive identities
(2.15) , the Jacobian is shown to be non-vanishing and give rise
to (3.3) and (3.5) ; this evaluation of Jacobian corresponds to
 super-imposing the covariant regularization on the generalized
Pauli-Villars regularization (2.5)} but the variation of
action, which corresponds to the right-hand side of (2.16), gives rise to
 the same anomaly factor as in (3.5) for $\Lambda\rightarrow\infty$.
\par
By recalling that $\gamma_{5}\phi_{n}(x)$ belongs to the eigenvalue
$-\lambda_{n}$ in (3.2) , the integration of (3.5) gives rise to

\begin{eqnarray}
n_{+}-n_{-}&=&\sum_{n}\int\phi_{n}(x)^{\dag}\gamma_{5}f(\lambda^{2}_{n}
/\Lambda^{2})\phi_{n}(x)d^{4}x\nonumber\\
&=&\frac{g^{2}}{32\pi^{2}}\int{Tr}\epsilon^{\mu\nu\alpha\beta}F_{\mu\nu}
F_{\alpha\beta}d^{4}x\nonumber\\
&=&\nu
\end{eqnarray}
where $n_{\pm}$ stand for the number of zero modes with $\gamma_{5}=\pm1$
,and $\nu$ is the Pontryagin index (or instanton number) \cite{12}.
Note that the replacement (3.16) is consistent with (3.17).
{}From (3.15) and (3.17), we obtain

\begin{equation}
d\mu\rightarrow{d}{\mu}exp\left[-i\alpha\nu\right]
\end{equation}
for the x-independent $\alpha$ in (3.14).
In (2.17), the variation of the action gives rise to the same phase
factor as (3.18) for $\Lambda\rightarrow\infty$.
Eqs.(3.14) and (3.18) show that
\begin{equation}
<\psi_{R}(x)\cdots\overline{\psi}_{R}(y)\cdots>=\int{d}\mu\left[\psi_{R}(x)
\cdots\overline{\psi}_{R}(y)\cdots\right]exp\left[\int{\cal L}d^{4}x\right]
\end{equation}
is non-vanishing only for the Green's functions which contain $\nu$ more
$\psi$ variables than $\overline{\psi}$ variables.
This gives the chirality selection rule and the fermion number non-
conservation.
Since the zero modes do not appear in the action (3.13), the path
integral (i.e.,  left-derivative) over Grassmann variables corresponding
to zero modes is completely consumed by $n_{+}\ \  \psi$-variables and $n_{-}
 \ \ \overline{\psi}$-variables appearing in Green's functions.
As a result, the Green's function and current operators do not contain
the (singular) inverse of zero eigenvalues any more, which could arise
from the action.
This procedure is thus consistent for anomaly-free gauge theory.
\par
When the gauge group contains anomaly, the Jacobian for the gauge
transformation (2.8) with gauge field kept fixed gives the covariant
anomaly (3.3) for the Noether current.
In this case, the variation of the partition function under the change of
 path integral variables has a definite meaning, but the partition
function itself is ill-defined since the anomalous gauge theory cannot be
 completely regularized by gauge invariant cut-off in terms of
$\lambda_{n}$:
To define the partition function, one needs to use a regulator which
explicitly breaks gauge invariance such as the $\underline{conventional}$
 Pauli-Villars regularization \cite{13}.

\section{Covariant Regularization of  Anomaly-free Theory}

\par
In view of the partial regularization (2.32) of the generalized Pauli-
Villars regularization, it is interesting to apply the fully
regularized expressions (3.1) for the practical calculations in anomaly-
free gauge theory.
We then enjoy much more freedom in choosing the regulator $f(x^{2})$
, simply because we do not require a Lagrangian-level implementation  of
$f(x^{2})$.
We illustrate this application by evaluating the vacuum polarization
tensor.
To be specific, we evaluate the vacuum polarization tensor for QED (in
Euclidean metric)
\begin{equation}
{\cal L}=\overline{\psi}i\gamma^{\mu}(\partial_{\mu}-ieA_{\mu})\psi-m
\overline{\psi}\psi
\end{equation}
by choosing a simple regulator
\begin{equation}
f(x^{2})=(\frac{\Lambda^{2}}{x^{2}+\Lambda^{2}})^{2}
\end{equation}
which is convenient for practical calculations and satisfies the condition
(2.31).
The final result can be readily extended to the chiral theory defined by
the first current in (3.1).
\par
We thus start with a regularized current

\begin{equation}
<\overline{\psi}(x)\gamma^{\mu}\psi(x)>_{cov}=\lim_{y{\rightarrow}x}Tr
\{\gamma^{\mu}\frac{1}{i\not{\!\!D}-m}(\frac{\Lambda^{2}}{\not{\!\!D}^{2}
+\Lambda^{2}})^{2}\delta(x-y)\}
\end{equation}
 with
\begin{equation}
\not{\!\!D}=\gamma^{\mu}(\partial_{\mu}-ieA_{\mu})
\end{equation}

\par
By expanding (4.3) in powers of $eA_{\mu}$ and retaining only the terms
linear in $eA_{\mu}$, one obtains
\begin{eqnarray}
\lefteqn{<\overline{\psi}(x)\gamma^{\mu}\psi(x)>_{cov}}\nonumber\\
&=&\lim_{y{\rightarrow}x}Tr\left\{\gamma^{\mu}\left[\frac{1}
{i\not{\!\partial}-m}
(-e\not{\!\!A})\frac{1}{i\not{\!\partial}-m}(\frac{\Lambda^{2}}{
\not{\!\partial}^{2}+\Lambda^{2}})^{2}\delta(x-y)\right]\right.\nonumber\\
&&+\gamma^{\mu}\frac{1}{i\not{\!\partial}-m}(\frac{\Lambda^{2}}{
\not{\!\partial}^{2}+\Lambda^{2}})^{2}(ie)\{(\partial_{\nu}A^{\nu})+
2A^{\nu}\partial_{\nu}+\frac{1}{4}\left[\gamma^{\alpha},
\gamma^{\nu}\right]F_{\alpha
\nu}\}\nonumber\\
\ \ &&\ \ \ \ \ \ \ \
\times(\frac{1}{\not{\!\partial}^{2}+\Lambda^{2}})\delta(x-y)\nonumber\\
&&+\gamma^{\mu}\frac{1}{i\not{\!\partial}-m}(\frac{1}{
\not{\!\partial}^{2}+\Lambda^{2}})(ie)\{(\partial_{\nu}A^{\nu})+
2A^{\nu}\partial_{\nu}+\frac{1}{4}\left[\gamma^{\alpha},
\gamma^{\nu}\right]F_{\alpha
\nu}\}\nonumber\\
\ \ &&\ \ \ \ \ \ \ \
\left.\times(\frac{\Lambda^{2}}{\not{\!\partial}^{2}+\Lambda^{2}})^{2}
\delta(x-y)\right\}\nonumber\\
\end{eqnarray}
where we used

\begin{eqnarray}
\not{\!\!D}^{2}&=&\frac{1}{2}\{\gamma^{\mu},\gamma^{\nu}\}D_{\mu}D_{\nu}+
\frac{1}{2}\left[\gamma^{\mu},\gamma^{\nu}\right]D_{\mu}D_{\nu}\nonumber\\
&=&D_{\mu}D^{\mu}-\frac{ie}{4}\left[\gamma^{\mu},
\gamma^{\nu}\right]F_{\mu\nu}\nonumber\\
&=&\partial_{\mu}\partial^{\mu}-ie\left[\partial_{\mu}A^{\mu}+A^{\mu}\partial
_{\mu}\right]-e^{2}A_{\mu}A^{\mu}-\frac{ie}{4}\left[\gamma^{\mu},
\gamma^{\nu}\right]
F_{\mu\nu}\nonumber\\
&=&\partial_{\mu}\partial^{\mu}-ie\left[(\partial_{\mu}A^{\mu})
+2A^{\mu}\partial
_{\mu}\right]-e^{2}A_{\mu}A^{\mu}-\frac{ie}{4}\left[\gamma^{\mu},
\gamma^{\nu}\right]
F_{\mu\nu}\nonumber\\
\end{eqnarray}
with $F_{\mu\nu}=\partial_{\mu}A_{\nu}-\partial_{\nu}A_{\mu}$.
The derivative operator $\partial_{\mu}$ in (4.5) and (4.6), except the
one in $(\partial_{\mu}A^{\mu})$ and $F_{\mu\nu}$, acts on $\underline{
all}$ the x-variables standing on the right of it.
By taking the variational derivative of (4.5) with respect to $eA_{\nu}
(z)$, one finds a regularized expression of the vacuum polarization
tensor

\begin{eqnarray}
\lefteqn{\lim_{y{\rightarrow}x}Tr\left\{\gamma^{\mu}\frac{(-1)}
{i\not{\!\partial}-m}
\gamma^{\nu}\delta(x-z)(\frac{\Lambda^{2}}
{\not{\!\partial}^{2}+\Lambda^{2}})^{2}
\delta(x-y) \right.} \nonumber\\
&&+\gamma^{\mu}\frac{1}{i\not{\!\partial}-m}
(\frac{\Lambda^{2}}{\not{\!\partial}^{2}+\Lambda^{2}})^{2}(i)
\left[(\partial^{\nu}
\delta(x-z))+2\delta(x-z)\partial^{\nu}\frac{}{}\right.\nonumber\\
&&\left.+\frac{1}{2}\left[\gamma^{\alpha},\gamma^{\nu}\right]
(\partial_{\alpha}\delta(x-z))\right]
\times(\frac{1}{\not{\!\partial}^{2}+\Lambda^{2}})\delta(x-y)\nonumber\\
&&+\gamma^{\mu}\frac{1}{i\not{\!\partial}-m}
(\frac{1}{\not{\!\partial}^{2}+\Lambda^{2}})(i)\left[(\partial^{\nu}
\delta(x-z))+2\delta(x-z)\partial^{\nu}\frac{}{}\right.\nonumber\\
&&\left.\left.+\frac{1}{2}\left[\gamma^{\alpha},\gamma^{\nu}\right]
(\partial_{\alpha}\delta(x-z))\right]
\times(\frac{\Lambda^{2}}{\not{\!\partial}^{2}+\Lambda^{2}})^{2}
\delta(x-y)\right\}\nonumber\\
\end{eqnarray}

\par
We now use (2.24) in (4.7), and we obtain the momentum representation
of the vacuum polarization tensor (see also (2.23))

\begin{eqnarray}
\Pi^{\mu\nu}(q)&=&\int\frac{d^{4}k}{(2\pi)^{4}}Tr\left\{\gamma^{\mu}\frac{(-1)}
{\not{\!k}+\not{\!q}-m}\gamma^{\nu}\frac{1}{\not{\!k}-m}
(\frac{\Lambda^{2}}{-k^{2}+\Lambda^{2}})^{2}\right.\nonumber\\
& &+\gamma^{\mu}\frac{1}{\not{\!k}+\not{\!q}-m}(\frac{\Lambda^{2}}{-(k+q)^{2}+
\Lambda^{2}})^{2}\left[q^{\nu}+2k^{\nu}
+\frac{1}{2}\left[\gamma^{\alpha},\gamma^{\nu}
\right]q_{\alpha}\right]\nonumber\\
&&\ \ \ \ \ \ \ \ \times(\frac{1}{-k^{2}+\Lambda^{2}})\nonumber\\
& &+\gamma^{\mu}\frac{1}{\not{\!k}+\not{\!q}-m}(\frac{1}{-(k+q)^{2}+
\Lambda^{2}})\left[q^{\nu}+2k^{\nu}+\frac{1}{2}
\left[\gamma^{\alpha},\gamma^{\nu}
\right]q_{\alpha}\right]\nonumber\\
&&\ \ \ \ \ \ \ \
\left.\times(\frac{\Lambda^{2}}{-k^{2}+\Lambda^{2}})^{2}\right\}.
\end{eqnarray}
The first term in (4.8) stands for the naive momentum cut-off by a form
factor, which generally spoils gauge invariance.
The remaining two terms in (4.8) recover gauge invariance spoiled by
the first term.
\par
After the standard trace calculation and using the Feynman parameters,
the first term in (4.8) gives

\begin{eqnarray}
\lefteqn{(\frac{\Lambda^{4}}{4\pi^{2}})\int^{1}_{0}d\alpha\int^{1-\alpha}_{0}
d\beta\left\{-g^{\mu\nu}\frac{\beta}{-\alpha(1-\alpha)q^{2}+(1-\beta)m^{2}
+\beta
\Lambda^{2}} \right.} \nonumber\\
&&\ \ \
+\left[-m^{2}g^{\mu\nu}-\alpha(1-\alpha)g^{\mu\nu}q^{2}+2\alpha(1-\alpha)
q^{\mu}q^{\nu}\right]\nonumber\\
&&\ \ \ \ \ \ \left.\times\frac{\beta}{\left[-\alpha(1-\alpha)q^{2}
+(1-\beta)m^{2}+\beta\Lambda^{2}\right]
^{2}}\right\}\nonumber\\
&\rightarrow&(\frac{1}{4\pi^{2}})\int^{1}_{0}d\alpha2\alpha(1-\alpha)
(q^{\mu}q^{\nu}-g^{\mu\nu}q^{2})ln\left[\frac{\Lambda^{2}}
{-\alpha(1-\alpha)q^{2}
+m^{2}}\right]\nonumber\\
& &\ \ \ \ \ \
+(\frac{1}{4\pi^{2}})\left[\frac{-1}{2}(\Lambda^{2}-m^{2})g^{\mu\nu}+
\frac{1}{6}g^{\mu\nu}q^{2}\right.\nonumber\\
&& \ \ \ \ \ \ \ \left.-\frac{1}{3}q^{\mu}q^{\nu}-\frac{5}{18}
(q^{\mu}q^{\nu}-g^{\mu\nu}q^{2})\right]
\end{eqnarray}
for $\Lambda\rightarrow{large}$.
Similarly, the second and third terms in (4.8) together give

\begin{eqnarray}
\lefteqn{(\frac{\Lambda^{4}}{4\pi^{2}})\int^{1}_{0}d\alpha\int^{1-\alpha}_{0}
d{\beta}\left\{g^{\mu\nu}\frac{1-\beta}{-\alpha(1-\alpha)q^{2}+{\beta}m^{2}
+(1-\beta)\Lambda^{2}} \right. }\nonumber\\
& &\ \ \ \ \
-\left[\alpha(2\alpha-1)q^{\mu}q^{\nu}+\alpha(g^{\mu\nu}q^{2}-q^{\mu}q^{\nu}
)\right]\nonumber\\
& &\ \ \ \ \ \ \ \ \
\left.\times\frac{1-\beta}{\left[-\alpha(1-\alpha)q^{2}+{\beta}m^{2}
+(1-\beta)\Lambda^{2}\right]
^{2}}\right\}\nonumber\\
&\rightarrow&(\frac{1}{4\pi^{2}})\left[ \frac{1}{2}
(\Lambda^{2}-m^{2})g^{\mu\nu}+
\frac{5}{36}g^{\mu\nu}q^{2} \right. \nonumber\\
&& \ \  \ \ \ \ \ \left. +\frac{1}{36}q^{\mu}q^{\nu}+\frac{1}{4}
(q^{\mu}q^{\nu}-g^{\mu\nu}q^{2})\right]
\end{eqnarray}
for $\Lambda\rightarrow{large}$.
These two expressions in (4.9) and (4.10) put together finally give rise
to the familiar gauge invariant result

\begin{equation}
(\frac{1}{4\pi^{2}})(q^{\mu}q^{\nu}-g^{\mu\nu}q^{2})\left\{\int^{1}_{0}d\alpha
2\alpha(1-\alpha)ln\left[\frac{\Lambda^{2}}{-\alpha(1-\alpha)q^{2}+m^{2}}
\right]
-\frac{1}{3}\right\}.
\end{equation}
The result for the chiral gauge theory (2.1) is obtained from (4.11) by
setting $m=0$ and multiplying it by $\frac{1}{2}TrT^{a}T^{b}$.
\par
The covariant regularization scheme thus gives rise to a gauge
invariant result on the basis of well-regularized finite calculations.
It is important that we always stay in d=4 dimensional space-time in this
 calculation.
This property is crucial for a reliable treatment of the anomaly.
The coefficient of $ln\Lambda^{2}$ in (4.11), which is related to the
renormalization group $\beta$-function, is independent of the choice of $
f(x^{2})$ in (4.2). For example, one can confirm that

\begin{equation}
f(x^{2})=(\frac{\Lambda^{2}}{x^{2}+\Lambda^{2}})^{n}\ ,\ n\geq2
\end{equation}
gives the same numerical coefficient of $ln\Lambda^{2}$ by dividing (4.9)
 and (4.10) by $\Lambda^{4}$ and taking suitable derivatives with respect
 to $\Lambda^{2}$.
The finite term, -1/3, in (4.11) depends on the specific regulator ; this
 is not a drawback since the finite term is uniquely fixed by the
renormalization condition in renormalizable theory.
The regulator independence of the coefficient of $ln\Lambda^{2}$ is also
expected from the fact that the Weyl anomaly in (3.6) (and related $
\beta$-function)is independent of $f(x^{2})$.
\par
The present covariant regularization can be readily applied to the
calculations of higher point functions and to practical calculations in
chiral gauge theory such as the Weinberg-Salam theory ; the covariant
 regularization can handle gauge anomalies in a reliable way , and thus one
can treat lepton and quark sectors separately without grouping
the fermions into a multiplet of SO(10).
The Higgs coupling, which mixes left-and right-handed components, is
readily handled by the present method as is explained in \cite{9}.

\section{Discussion and Conclusion}

Motivated by the interesting suggestion of generalized Pauli-Villars
regularization, we re-examined the regularization and anomalies in
gauge theory.
The generalized Pauli-Villars regularization as reformulated as a
 regularization of component operators in this
paper will perhaps make the covariant regularization, which has been
known for some time, more acceptable ;  the Lagrangian level
realization of the covariant regularization for anomaly-free gauge
theory however remains as an open question.
The covariant regularization spoils the Bose symmetry for anomalous gauge
 theory, but it preserves the Bose symmetry as well as gauge invariance
 for anomaly-free gauge theory.
\par
Our analysis here is confined to one-loop level calculations, though certain
non-perturbative aspects such as instantons are also involved.
As for multi-loop diagrams, the higher derivative regularization in
the sector of gauge fields \cite{14}, for example, can render all the
multi-loop diagrams finite.
The one-loop diagrams which include only the gauge fields cannot
 be regularized by the higher derivative regularization but they can be
covariantly regularized if one uses the covariant background gauge
technique\cite{15}.
In fact, we recently illustrated a simple non-diagramatic calculation
of one-loop $\beta$-function of QCD by using the method of covariant
anomaly\cite{16}.
\par
The generalized Pauli-Villars regularization is also known to have
interesting implications on lattice gauge theory \cite{17}, but its
analysis is beyond the scope of the present paper.
\par
In conclusion, we have shown that the basic mechanism of generalized
Pauli-Villars regularization of continuum theory is made transparent if
one looks at it from the view point of a regularization of composite
current operators ; by this way , one can readily compare the generalized
Pauli-Villars regularization with the covariant regularization of chiral
 gauge theory.
The covariant regularization scheme, which is quite flexible, has been
also shown to be useful in practical calculations.

\section*{Appendix}
\setcounter{equation}{0}
\setcounter{section}{0}
\renewcommand{\theequation}{A\arabic{equation}}
\renewcommand{\thesubsection}{A\arabic{subsection}}

\par
For the sake of completeness, we here quote the proof of f(x)-
independence of (3.3) and (3.5)\cite{6}.
The calculation of (3.5), for example, proceeds as

\begin{eqnarray}
\lefteqn{iTr\int\frac{d^{4}k}{(2\pi)^{4}}e^{-ikx}\gamma_{5}f(\not{\!\!D}^{2}/
\Lambda^{2})e^{ikx}}\nonumber\\
&=&iTr\int\frac{d^{4}k}{(2\pi)^{4}}\gamma_{5}f(\frac{(ik_{\mu}+D_{\mu})
(ik^{\mu}+D^{\mu})-\frac{ig}{4}\left[\gamma^{\mu},\gamma^{\nu}\right]
F_{\mu\nu}}{
\Lambda^{2}})\nonumber\\
&=&i\Lambda^{4}Tr\int\frac{d^{4}k}{(2\pi)^{4}}\gamma_{5}f(
-k_{\mu}k^{\mu}
+\frac{2ik^{\mu}D_{\mu}}{\Lambda}+\frac{D^{\mu}D_{\mu}}
{\Lambda^{2}}\nonumber\\
&&\ \ \ \ \ \ \ \
-\frac{ig}{4\Lambda^{2}}\left[\gamma^{\mu},\gamma^{\nu}\right]F_{\mu\nu})
\end{eqnarray}
where we used

\begin{eqnarray}
\not{\!\!D}^{2}&=&\frac{1}{2}\{\gamma^{\mu},\gamma^{\nu}\}D_{\mu}D_{\nu}
+\frac{1}{2}\left[\gamma^{\mu},\gamma^{\nu}\right]D_{\mu}D_{\nu}\nonumber\\
&=&D_{\mu}D^{\mu}-\frac{ig}{4}\left[\gamma^{\mu},\gamma^{\nu}\right]F_{\mu\nu}
\end{eqnarray}
and re-scaled the variable $k_{\mu}\rightarrow\Lambda{k}_{\mu}$.
We next expand the quantity involving f(x) in (A.1) around $x=-k_{\mu}
k^{\mu}=|k^{2}|$ as

\begin{eqnarray}
\lefteqn{f(-k_{\mu}k^{\mu}+\frac{2ik^{\mu}D_{\mu}}{\Lambda}
+\frac{D^{\mu}D_{\mu}}
{\Lambda^{2}}-\frac{ig}{4\Lambda^{2}}\left[\gamma^{\mu},\gamma^{\nu}\right]
F_{\mu\nu})}\nonumber\\
&=&f(-k_{\mu}k^{\mu})+f'(-k_{\mu}k^{\mu})\{\frac{2ik^{\mu}D_{\mu}}
{\Lambda}+\frac{D^{\mu}D_{\mu}}{\Lambda^{2}}-\frac{ig}{4\Lambda^{2}}
\left[\gamma^{\mu},\gamma^{\nu}\right]F_{\mu\nu}\}\nonumber\\
&&+\frac{1}{2!}f''(-k_{\mu}k^{\mu})\{\frac{2ik^{\mu}D_{\mu}}
{\Lambda}+\frac{D^{\mu}D_{\mu}}{\Lambda^{2}}-\frac{ig}{4\Lambda^{2}}
\left[\gamma^{\mu},\gamma^{\nu}\right]F_{\mu\nu}\}^{2}+...
\end{eqnarray}
When $\Lambda\rightarrow\infty$, only the terms of order $1/\Lambda^{4}$
or larger in (A.3) survive in (A.1).
Moreover, the trace $Tr(\gamma_{5}...)$ is non-vanishing only for the
terms with more than four $\gamma$-matrices.
The only term that satisfies these two conditions is the third term in (A.
3) with $(\left[\gamma^{\mu},\gamma^{\nu}\right]F_{\mu\nu})^{2}$.
 The calculation of
 (A.1) thus becomes

\begin{eqnarray}
\lefteqn{iTr\int\frac{d^{4}k}{(2\pi)^{4}}\gamma_{5}\frac{1}{2!}f''(-k_{\mu}k^{
\mu})\{\frac{-ig}{4}\left[\gamma^{\mu},\gamma^{\nu}\right]
F_{\mu\nu}\}^{2}}\nonumber\\
&=&iTr\gamma_{5}\frac{1}{2}\{\frac{-ig}{4}\left[\gamma^{\mu},
\gamma^{\nu}\right]
F_{\mu\nu}\}^{2}
\frac{1}{16\pi^{2}}\int^{\infty}_{0}dxxf''(x)\nonumber\\
&=&(\frac{ig^{2}}{32\pi^{2}})Tr\epsilon^{\mu\nu\alpha\beta}F_{\mu\nu}
F_{\alpha\beta}
\end{eqnarray}
after taking the trace over $\gamma$-matrices: We here used $d^{4}k=
\pi^{2}|k^{2}|d|k^{2}|$, and

\begin{eqnarray}
\int^{\infty}_{0}dx\ xf''(x)&=&xf'(x)|^{\infty}_{0}-\int^{\infty}_{0}
dxf'(x)\nonumber\\
&=&-f(x)|^{\infty}_{0}=f(0)=1
\end{eqnarray}
by noting the conditions (2.31) including $xf'(x)\rightarrow0$ for
 $x\rightarrow\infty$. The result (A.4) is thus independent of the
regulator f(x) ; the convenient choice of f(x) for practical
calculations is (4.12) or $f(x)=exp\left[-x\right] $.
\par
The above analysis is also applicable to (3.3).
It is known that a similar analysis holds for (3.6)\cite{8}.

{\bf Note Added}
\par
After submitting the present paper , the works by Narayanan and Neuberger
\cite{18} on the Kaplan's formulation came to my attention. These authors
analyze two-dimensional gauge anomaly , in particular consistent from of
anomaly , from a view point of 2+1 dimensional theory. This calculational
scheme is apparently different from the generalized Pauli-Villars
regularization in (2.5) , which is not applicable to the evaluation of gauge
anomaly.


\begin{thebibliography}{1}

\bibitem{1}
S.A.Frolov and A.A.Slavnov, Phys. Lett.{ \bf B309 }(1993)344

\bibitem{2}
R.Narayanan and H.Neuberger, Phys. Lett.{\bf  B301}(1993)62

\bibitem{3}
S.Aoki and Y.Kikukawa, Mod. Phys. Lett. {\bf A8}(1993)3517

\bibitem{4}
G.'t Hooft, Phys. Rev. Lett. {\bf 37}(1976)8 ; \\
Phys. Rev. {\bf D14}(1976)172 ; {\bf D18}(1978)2199(E)

\bibitem{5}
S.Adler, Phys.Rev. {\bf 177}(1969)2426\\
J.S.Bell and R.Jackiw, Nuovo Cim, {\bf A60}(1969)47

\bibitem{6}
K.Fujikawa, Phys. Rev. {\bf D21}(1980)2848; {\bf 22}(1980)1499(E);\\
Phys. Rev. Lett. {\bf 42}(1979)1195

\bibitem{7}
L.Alvarez-Gaum$\acute{e}$ and E.Witten, Nucl. Phys. {\bf B234}(1983)269

\bibitem{8}
K.Fujikawa, Phys. Rev. Lett. {\bf 44}(1980)1733;\\
Phys. Rev. {\bf D23}(1981)2262

\bibitem{9}
K.Fujikawa, Phys. Rev. {\bf D29}(1984)285

\bibitem{10}
W.A.Bardeen and B.Zumino, Nucl. Phys. {\bf B244}(1984)421

\bibitem{11}
J.Wess and B.Zumino, Phys. Lett. {\bf 37B}(1971)95

\bibitem{12}
M.Atiyah and I.Singer, Ann. Math. {\bf 87}(1968)484

\bibitem{13}
W.A.Bardeen, Phys. Rev. {\bf 184}(1969)1848\\
H.Shinke and H.Suzuki, Mod. Phys. Lett. {\bf A40}(1993)3835

\bibitem{14}
L.D.Faddeev and A.A.Slavnov,
$\underline{Gauge Fields}$ (Benjamin / Cummings, 1980)

\bibitem{15}
G.'t Hooft, Nucl. Phys. {\bf B62}(1973)444

\bibitem{16}
K.Fujikawa, Phys. Rev. {\bf D48}(1993)3922

\bibitem{17}
D.B.Kaplan, Phys. Lett. {\bf B288}(1992)342

\bibitem{18}
R.Narayanan and H.Neuberger , Phys. Rev. Lett. {\bf 71}(1993)3251 ; \\
Nucl. Phys. {\bf B412}(1994)574
\end{thebibliography}
\end{document}